\documentclass[10pt]{iopart}
\usepackage{iopams}
\usepackage{graphicx}
\usepackage{multirow}
\usepackage{diagbox}
\usepackage{bm}
\usepackage{hyperref}

\begin{document}

\title[Acceptors in (Hg,Mn)Te quantum wells]{Role of magnetic doping in topological HgTe and application of the Gram-Schmidt method for computing impurity states in quantum wells}

\author{D Bugajewski$^1$ and T Dietl$^{2}$}
\address{$^1$ Faculty of Physics, University of Warsaw, Pasteura $5$, PL-$02093$ Warsaw, Poland}
\address{$^2$ International Research Centre MagTop, Institute of Physics,\\
Polish Academy of Sciences, Aleja Lotnikow 32/46, PL-02668 Warsaw, Poland}
\ead{d.bugajewski@student.uw.edu.pl, dietl@MagTop.ifpan.edu.pl}

\vspace{10pt}
\begin{indented}
\item[]20 September 2024
\end{indented}

\begin{abstract}
The quantum spin Hall effect in non-magnetic and Mn-doped HgTe quantum well is strongly affected by Kondo scattering of edge electrons by holes localized on
acceptors.
A generalized eigenvalue method is usually employed for determining impurity binding energies from the multiband Kohn-Luttinger Hamiltonians in bulk samples and semiconductor quantum structures.
Such an approach provides accurate values of the level positions but its applicability for determining the impurity localization radius can
be questioned. As an alternative method we propose here the Gram-Schmidt orthogonalization procedure allowing to employ the standard eigenvalue algorithms and, thus, to determine
both impurity level energies and the set of normalized eigenvectors. We apply this approach to singly-ionized acceptor states in HgTe
quantum wells and obtain impurity level energies and localization radiuses even for states degenerate with the continuum of band states.
Such information allows us to
assess the energy of bound magnetic polarons in quantum wells doped with magnetic ions. We determine the
polaron energies and discuss consequences of the resonant polaron formation on band transport in the bulk samples and quantum wells in the regimes of quantum Hall effects.
\end{abstract}

\vspace{2pc}

\ioptwocol
{\let\newpage\relax\maketitle}

\section{Introduction}
According to the Mott criterion $N^{1/3}a^*=0.26$ \cite{mott}, residual donors with the volume concentration $N$ down to $10^{14}$\,cm$^{-3}$ cannot bind electrons in narrow gap zinc-blende semiconductors, where the donor localization radius $a^*$ is large. In contrast, due to the presence of the heavy hole band, the expected values of $a^*$ allow for holes' trapping by acceptors with concentrations up to $10^{17}$\,cm$^{-3}$.

The role of acceptor states is particularly
intriguing in zero-gap regime in HgTe as well as in Hg$_{1-x}$Cd$_x$Te, Hg$_{1-x}$Mn$_x$Te, and related mixed crystals with sufficiently low $x$ values, in which acceptor states reside above the top of the valence band and form resonant states in the {\em conduction} band \cite{Gelmont:1972_JETF,Bastard:1973_PLA}.
It was initially argued that a conductance minimum, observed in
zero-gap compounds around 30\,K, resulted from resonant scattering of electrons once the Fermi energy reaches the acceptor level \cite{Liu:1976_SSC}. However, detailed transport theory described the conductance anomalies in terms of interband optical phonon scattering without any adjustable parameters \cite{Dubowski:1981_JPCS}. However, as concentrations of residual acceptors are typically larger than those of donors, the acceptors states pin the Fermi level at low temperatures \cite{Szlenk:1979_pssb,Szlenk:1979_pssb_b,Brandt:1982_JETP,Sawicki:1983_Pr}. At the same time, the concept of the Efros-Shklovskii Coulomb gap and the associated charge ordering \cite{Bello:1981_JETF,Wilamowski:1990_SSC} explained qualitatively (i) decoupling of the resonant levels and conduction band states and (ii) the surprising {\em enhancement} of electron mobility observed at low temperatures \cite{Brandt:1982_JETP,Sawicki:1983_Pr}. An additional contribution from low-mobility electrons deduced from the mobility spectrum was assigned to hopping conductance in the acceptor band \cite{Szlenk:1979_pssb} but the role of the camel band structure of the valence band, brought about by the lack of inversion symmetry in the zinc-blende structure, is to be clarified.

Several arguments point to a substantial concentration of residual acceptors also
in the case of HgTe or (Hg,Mn)Te quantum wells (QWs): (i) undoped samples are $p$-type and modulation doping by I \cite{Konig:2007_S,Bendias:2018_NL} or In \cite{Grabecki:2013_PRB,Yahniuk:2019_QM} donors is necessary to shift the Fermi level to the conduction
band; (ii) despite a relatively low effective mass and large dielectric constance, electron mobility is smaller
than in CdTe and (Cd,Mn)Te QWs doped with I \cite{Piot:2010_PRB,Betthausen:2014_PRB}; (iii) the gate voltage necessary to shift
the Fermi energy from the conduction to the valence band points to the in-gap level concentration
of the order $10^{11}$\,cm$^{-2}$ \cite{Dietl:2023_PRL,Dietl:2023_PRB}, the value consistent with the magnitude of electron mobility. According to the theory \cite{Dietl:2023_PRL,Dietl:2023_PRB}, the acceptor levels are degenerate with topological edge states or with the {\em valence} part of the Dirac cones for the QW thickness $d_{\mathrm{QW}}$ corresponding to the topological phase transition, i.e., to the zero-gap regime.


The acceptor band model \cite{Dietl:2023_PRL,Dietl:2023_PRB} explains several
surprising observations near the topological phase transition, including: (i) a slow increase of hole densities with the gate voltage \cite{Bendias:2018_NL,Yahniuk:2021_arXiv}; (ii) the presence of an unusually wide quantum Hall plateau in the hole transport regime \cite{Konig:2007_S,Yahniuk:2021_arXiv,Yahniuk:2019_QM,Shamim:2020_SA}, and (iii) higher hole mobility values compared to the electron case \cite{Bendias:2018_NL,Yahniuk:2021_arXiv}.
Similarly to the bulk case, a contribution to the conductance can come from hopping in the acceptor band or transport over side valence band maxima resulting from both inversion asymmetry and coupling between QW subbands \cite{Novik:2005_PRB}.

The acceptor band plays also a predominant role in the topological regime. Here, a small energetic distance between edge electrons and acceptor holes results in a giant antiferromagnetic coupling between spins of electrons and holes localized near the edge \cite{Dietl:2023_PRL,Dietl:2023_PRB,Altshuler:2013_PRL} making that Kondo temperature $T_{\mbox{K}}$ is as high as a couple of Kelvins despite a low density of states in the electron edge cones \cite{Dietl:2023_PRL,Dietl:2023_PRB}. Under these conditions, hole-induced backscattering between helical channels affects the edge conductance in the quantum spin Hall regime according to
\begin{equation}
G = (2e^2/h)(1+ L_x/L_{\mathrm{p}})^{-1},
\label{eq:G}
\end{equation}
where $L_x$ is the edge length and $L_{\mathrm{p}}$ is the topological protection length which in the Kondo regime is given by \cite{Dietl:2023_PRL,Dietl:2023_PRB,Maciejko:2009_PRL,Tanaka:2011_PRL,Vayrynen:2016_PRB},
\begin{equation}
 L_{\mathrm{p}}^{-1}(T) = \sum_ir^{(i)}f(T/T_{\mathrm{K}}^{(i)})/L_x.
 \label{eq:Lp}
\end{equation}
Here the summation is over all QW holes bound to acceptors for a given gate voltage $V_{\mbox{g}}$; $r$ is a ratio of the Kondo scattering rates for transitions that do not conserve and conserve the total spin of the electron and hole system, and contains a temperature dependent correction resulting from electron-electron coupling; $f(x)$ is a function obtained by the numerical renormalization group approach \cite{Costi:2009_PRL} that is more accurate than the original Nagaoka-Suhl
resummation result, $f(x) = a/[a + \ln^2(T/T_{\mathrm{K}})]$, where $a = 3\pi^2/4$. In either case, $f(0) = 0;\, f (1)\simeq 1$, and $f(10) \simeq 0.6$. As $T_{\mathrm{K}}$ decays
exponentially away from the edge region, the formula for $L_{\mathrm{p}}$ means that the electron mean free path is essentially equal to an average distance between acceptor holes
located along the electron trajectory reduced by the factor $r$, typically of the order of $10^{-2}$.

 Particulary striking results have been obtained by doping of HgTe QWs by Mn impurities. Counterintuitive, such doping enhances hole mobility to almost $10^6$\,cm$^2$/Vs \cite{Shamim:2020_SA} and increases substantially the topological protection length \cite{Shamim:2021_NC}. Those results were explained by the formation of acceptor bound magnetic polarons, and the resulting weakening of Kondo scattering of itinerant electrons by acceptor holes. A quantification of those effects requires detailed information on the acceptor wave function.

A theoretical analysis of the acceptor states in semiconductor quantum wells is usually based on the effective mass approximation taking into account heavy and light hole bands in the axial approximation \cite{Fraizzoli:1991_PRB,Kozlov:2019_S}. Those approaches were recently generalized \cite{Dietl:2023_PRL,Dietl:2023_PRB} by employing the eight band Luttinger-Kohn model with appropriate boundary conditions at the interfaces \cite{Novik:2005_PRB}. As in previous works \cite{Fraizzoli:1991_PRB,Buczko:1992_PRB}, the in-plane envelope consisted of a set of exponentially decaying functions with the radii forming a geometrical series \cite{Dietl:2023_PRL,Dietl:2023_PRB}. As such a set is not orthonormal, a generalized eigenvalue solver served to determine the acceptor levels. At the same time, the participation number was employed to estimate magnitudes of the in-plane localization radius $a^*$.

In this paper, we use the Gram-Schmidt orthogonalization procedure (see e.g. \cite{Liu:2022_B}) to construct a new set of orthononormal radial base envelope functions, which then allows us to use a standard
eigenvalue solver to obtain both eigenvalues and orthonormal eigenvectors. Not surprisingly, those two methods provide numerically identical level energies but magnitudes of $a^*$ differ. In Sec.~\ref{sec:math}, the derivation of the mathematical method is outlined, while in Sec.~\ref{sec:res}, we present and analyze the obtained numerical results. Computations of energy levels and localization radii $a^*$ as a function of the acceptor charge $Z$ allow us to tell the band and resonant acceptor states. As shown in Sec.~\ref{sec:BMP}, the newly determined values of $a^*$ make it possible to evaluate hopping characteristics and the influence of bound magnetic polarons upon the Kondo effect and the Coulomb gap in (Hg,Mn)Te QWs.

\section{Acceptor level energies and in plane-localization radii}
\subsection{The method and its convergence}\label{sec:math}
In previous works \cite{Dietl:2023_PRL,Dietl:2023_PRB}, the eight-band Kohn-Luttinger Hamiltonian in the axial approximation, supplemented by the Coulomb potentials of the acceptor and its image charges, served to determine energy levels in a quantum well grown along the $z$ direction. Within that model, the electron wave function at given integer orbital momentum $m$ is
assumed to be a linear combination of the normalized Kohn-Luttinger amplitudes $u_j$ with appropriate envelope functions,
\begin{equation}
\psi_j(\rho,\phi,z)=f_j(\rho,\phi)h_j(z)u_j,
\end{equation}
where $u=[s_{1/2,1/2}$, $s_{1/2,-1/2}$, $p_{3/2,3/2}$, $p_{3/2,1/2}$, $p_{3/2,-1/2}$, $p_{3/2,-3/2}$, $p_{1/2,1/2}$, $p_{1/2,-1/2}]$ is the set of the normalized Kohn-Luttinger amplitudes for $s_z=\pm1/2$, $j_z =\pm3/2$ and $\pm1/2$, $h_j(z)$ are linear combinations of complex exponential functions
\begin{equation}
h_j^{(n)} =\exp(2\pi\mbox{i}n z/L_{\mathrm{z}})/\sqrt{L_z},
\end{equation}
where $n$ are integer numbers, $n\in\left\{-n_{\mathrm{max}},...,n_{\mathrm{max}}\right\}$; $L_{\mathrm{z}}$ is the total width of the structure consisting of two barriers and the quantum well, $L_{\mathrm{z}}= 2d_{\mathrm{b}} +d_{\mathrm{QW}}$, and
\begin{equation}
f_j^{(m_j)}(\rho,\phi)=\mbox{i}^{|m_j|}\exp(\mbox{i}m_j\phi)F_j(\rho)/(2\pi)^{1/2},
\end{equation}
where $m_1=m_4=m_7=m$, $m_2=m_5=m_8=m+1$, $m_3=m-1$, $m_6=m+2$ ($m_j$ -- integer orbital magnetic quantum numbers corresponding to the $z$ component of the orbital momentum; the total angular momentum is $m+1/2$ \cite{Dietl:2023_PRB}) and the functions $F_j(\rho)$ are linear combinations of exponential functions of the form
\begin{equation}\label{no}
F_j^{(l,m_j)} =N_{jl}\rho^{|m_j|}\exp(-\rho/a_l),
\end{equation}
with $N_{jl}$ being a normalization factor; $l\in\left\{1,...,l_{\mathrm{max}}\right\}$, and $a_l$ values are chosen as a geometric sequence.

As the functions $F_j^{(l,m_j)}(\rho)$ are not orthogonal, a generalized eigenvalue solver had to be employed, which provided $8(2n_{\mathrm{max}} +1)l_{\mathrm{max}}$ energies and, in general, not normalized
 coefficients $b_{j,n,l}$ describing the participation of particular eigenvectors $u_jh_j^{(n)}f_j^{(l)}$ in the electron wave function for a given $m$ and eigenvalue.

\begin{figure}
  \centering
  \includegraphics[width=\columnwidth]{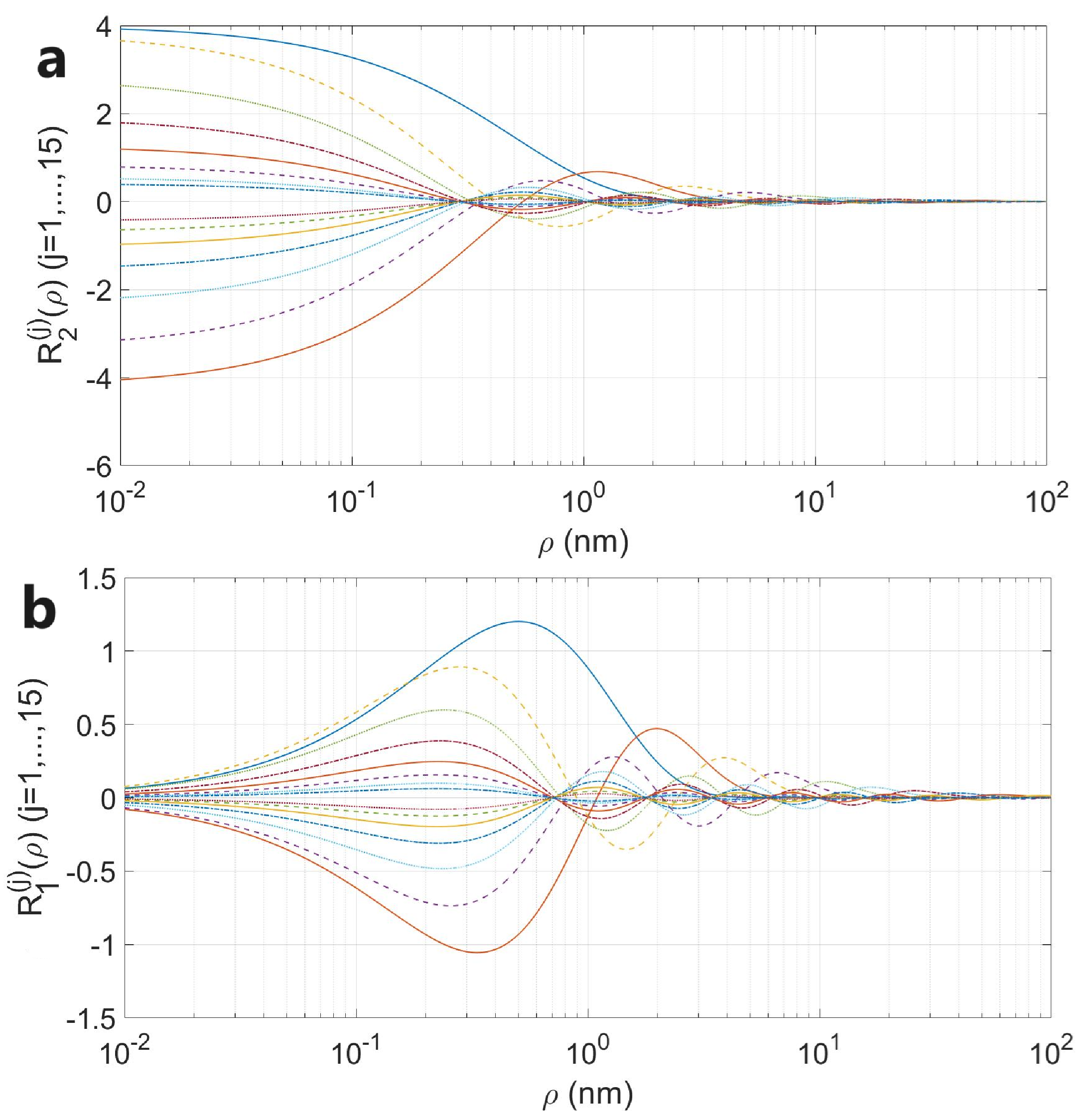}
  \caption{\label{fig1} The sets of orthonormal functions $R_l^{|m_j|}(\rho)$ obtained using the Gram-Schmidt orthogonalization procedure: (a) $m_j=0$ (b) $m_j=1$.}
\end{figure}

In the present paper, we replace the sets of $l$ functions contributing to $F_j(\rho)$ with an orthonormal base constructed by applying the Gram-Schmidt orthogonalization procedure \cite{Liu:2022_B},
that is, we take the envelope functions $f_j^{(m_j)}(\rho,\phi)$ as linear combinations
of functions given now by
\begin{eqnarray}
f_j^{(l,m_j)}(\rho,\phi)=\mbox{i}^{|m_j|}\exp(im_j\phi)R_j^{(l)}(\rho)/(2\pi)^{1/2},
\end{eqnarray}
where
\begin{eqnarray}\label{ort}
R_j^{(l,m_j)}(\rho)=\rho^{|m_j|}\sum\limits_{k=1}^lw_{|m_j|}(k,l)\exp(-\rho/a_l),
\end{eqnarray}
and the coefficients $w_{|m_j|}(k,l)$ are to be determined by the orthogonalization procedure. Figure \ref{fig1} presents the obtained orthonormal basis for some exemplary values of $m_j$. In the new base of envelope functions, the formula for the in-plane localization radius becomes
\begin{eqnarray}
a^*=\ \\ \nonumber \frac{\sum_{j,n,l}|b_{j,n,l}|^2}{\left| \sum_{j,n,j',n',l_{1-4}}b_{j,n,l_1}b^*_{j,n,l_2}b_{j',n',l_3}b^*_{j',n',l_4}g_{l_{1-4}}^{(j,j')} \right| ^{1/2}}, 
\end{eqnarray}
where $l_{1-4}:=\left\{l_1,...,l_4\right\}$ and
\begin{eqnarray}
g^{(j,j')}_{l_{1-4}}= \ \\ \nonumber
\sum_{k_{1-4}=1}^{l_{1-4}}\left(w_{|m_j|}(k_1,l_1)w_{|m_j|}(k_2,l_2)w_{|m_j'|}(k_3,l_3)\right.\ \\ \nonumber
\left.w_{|m_j'|}(k_4,l_4)\frac{(2|m_j|+2|m_{j'}|+1)!}{(a_{k_1}^{-1}+a_{k_2}^{-1}+a_{k_3}^{-1}+a_{k_4}^{-1})^{2|m_j|+2|m_{j'}|+2}}\right).
\end{eqnarray}
It is worth mentioning that $\sum_{j,n,l}|b_{j,n,l}|^2 =1$, as in this case a standard eigenvalue solver, providing orthonormal eigenvectors, has been employed.

One can also estimate the inverse participation ratio $z^*$ of the electron's wave function along the $z$ axis using the formula
\begin{eqnarray}
z^*=\ \\ \nonumber \frac{L_z\sum_{j,n,l}|b_{j,n,l}|^2}{\left| \sum_{j,j',l,l',n_{1-4}}\delta_nb_{j,n_1,l}b^*_{j,n_2,l}b_{j',n_3,l'}b^*_{j',n_4,l'} \right|},
\end{eqnarray}
where $n_{1-4}:=\left\{n_1,...,n_4\right\}$, $\delta_n=1$ if $n_1+n_3=n_2+n_4$ and $\delta_n=0$ otherwise. 
Like in the formula for $a^*$, $\sum_{j,n,l}|b_{j,n,l}|^2 =1$ in this case.

\subsection{Numerical results}\label{sec:res}
We apply our diagonalization methods to HgTe quantum wells employing the same material parameters as before \cite{Dietl:2023_PRL,Dietl:2023_PRB}, including the barrier thickness $d_{\mathrm{b}}= 30 $\,nm and
quantum well thicknesses $d_{\mathrm{QW}} = 8$ and 6\,nm, for which the topological band gap is about 20 and 0\,meV . We assume zero energy at the top of the bulk HgTe $\Gamma_8$ valence band. We consider the acceptor impurity localized in the QW centre.

\begin{figure}
  \centering
  \includegraphics[width=\columnwidth]{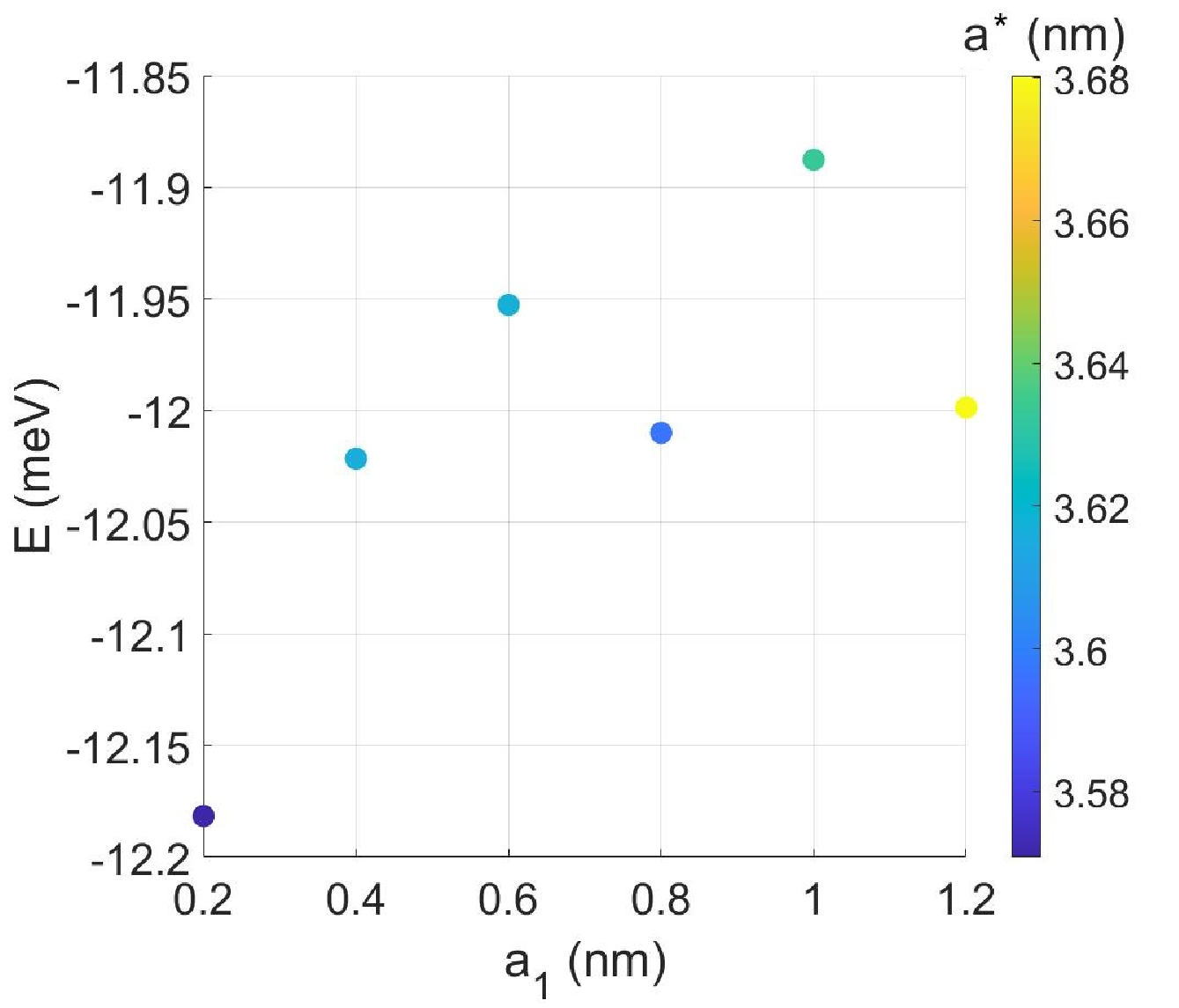}
  \caption{\label{fig2} The relationship between the values of $E$ and $a^*$ obtained for different values of $a_1$. The values differ only by approximately $2-3\%$, so the results are assumed to be independent of the choice of $a_1$.}
\end{figure}

We find that although the accuracy of calculations improves with the increase of $l_{\mathrm{max}}$, the numerical errors resulting from orthogonalization grow as well (see Table \ref{tabgramschmidt}) -- that is why we consider $l_{\mathrm{max}}=10$ as the optimal value. For example, for $d_{\mathrm{QW}} = 8$\,nm, $a_1=0.6\,\mathrm{nm}$, the common ratio $r = a_{l+1}/a_l =1.5$, and $n_{\mathrm{max}}=25$, we obtain $E=-11.95\,\mathrm{meV}$ and $a^*=3.617$\,nm for $l_{\mathrm{max}}=10$, while $E=-11.96\,\mathrm{meV}$ and $a^*=3.601\,\mathrm{nm}$ for $l_{\mathrm{max}}=15$ - these values are practically identical; more precisely, the differences resulting from change of $l_{\mathrm{max}}$ are smaller than the differences resulting from changing the value of $a_1$, and the numerical errors resulting from the Gram-Schmidt orthogonalization are significantly smaller for $l_{\mathrm{max}}=10$.
\begin{table} [h!]
\caption{\label{tabgramschmidt} Testing the numerical accuracy of the Gram-Schmidt procedure. As an example, we present the values $\langle f_0^{(l_1,0)}|f_0^{(l_2,0)}\rangle$ for $1\leq l_{1,2}\leq 15$. For $l_{1,2}\leq11$ (which corresponds to $l_{\mathrm{max}}=10$), $\langle f_0^{(l_1,0)}|f_0^{(l_2,0)}\rangle=\delta_{ij}$ as expected with respect to 4 decimal places. For $l>11$, bigger discrepancies start arising as a result of gradual overlapping of previous numerical errors.}
\vspace{4pt}
\scriptsize
\centering
  \begin{tabular}{c|ccccccc}
        \diagbox{$l_1$}{$l_2$} & \textbf{1} & $\ldots$ & \textbf{11} & \textbf{12} & \textbf{13} & \textbf{14} \\ \hline
        \textbf{1} & 1.0000 &  & 0.0000 & 0.0000 & 0.0000 & 0.0000   \\
        $\ldots$ &  &  &  &  &   &  &    \\
        \textbf{11} &  0.0000 &   & 1.0000 & 0.0001 & -0.0003 & 0.0005   \\
        \textbf{12} & 0.0000  &    & 0.0001  & 1.0000 & -0.0011 & 0.0034  \\
        \textbf{13} & 0.0000  &    & -0.0003 & -0.0011  & 1.0000 & 0.0146   \\
        \textbf{14} & 0.0000  &    & 0.0005 & 0.0034 & 0.0146 &  1.0000   \\

    \end{tabular}
\end{table}\\
Moreover, the results are stable when changing $a_1$ and $n_{\mathrm{max}}$. For instance, Fig.~\ref{fig2} presents values of $E$ and $a^*$ calculated for $j_{\mathrm{max}}=10$, $n_{\mathrm{max}}=25$, $m=0$, $d_{\mathrm{QW}}=8\,\mathrm{nm}$ and different values of $a_1$. The computational efficiency of the proposed method is similar to the standard scenario with non-orthogonal functions. The largest number of operations takes calculating $a^*$ and $z^*$ for already obtained eigenvectors -- however, if necessary, one can avoid this limitation by, e.g., first performing calculations with reduced accuracy to approximately locate the acceptor level, and then repeat the procedure with full accuracy but now counting the localization parameters for the identified level only. The method is also successful for larger values of $Z$ and $d_{\mathrm{QW}}$ than the ones investigated in this work, even if we decrease the number of functions $R_j^{(l,m_j)}$ taken into account: for example, we obtain (all energies $E$ in meV and $d_{\mathrm{QW}},a^*$ in nm, $n_{\mathrm{max}}=25$, $l_{\mathrm{max}}=7$) $E=-4.381$, $a^*=4.007$ ($d_{\mathrm{QW}}=10$, $Z=1$), $E=-0.001$, $a^*=4.451$ ($d_{\mathrm{QW}}=12$, $Z=1$), $E=2.616$, $a^*=4.804$ ($d_{\mathrm{QW}}=14$, $Z=1$), $E=-7.047$, $a^*=2.612$ ($d_{\mathrm{QW}}=6$, $Z=1.5$), $E=15.126$, $a^*=2.254$ ($d_{\mathrm{QW}}=6$, $Z=2$). Let us also emphasize that the orthogonalization procedure itself does not depend on $d_{\mathrm{QW}}$ or $Z$ [see formula (\ref{no})] -- the impact of these parameters is analogous to the non-orthogonal case.

\begin{figure}

  \centering
  \includegraphics[width=\columnwidth]{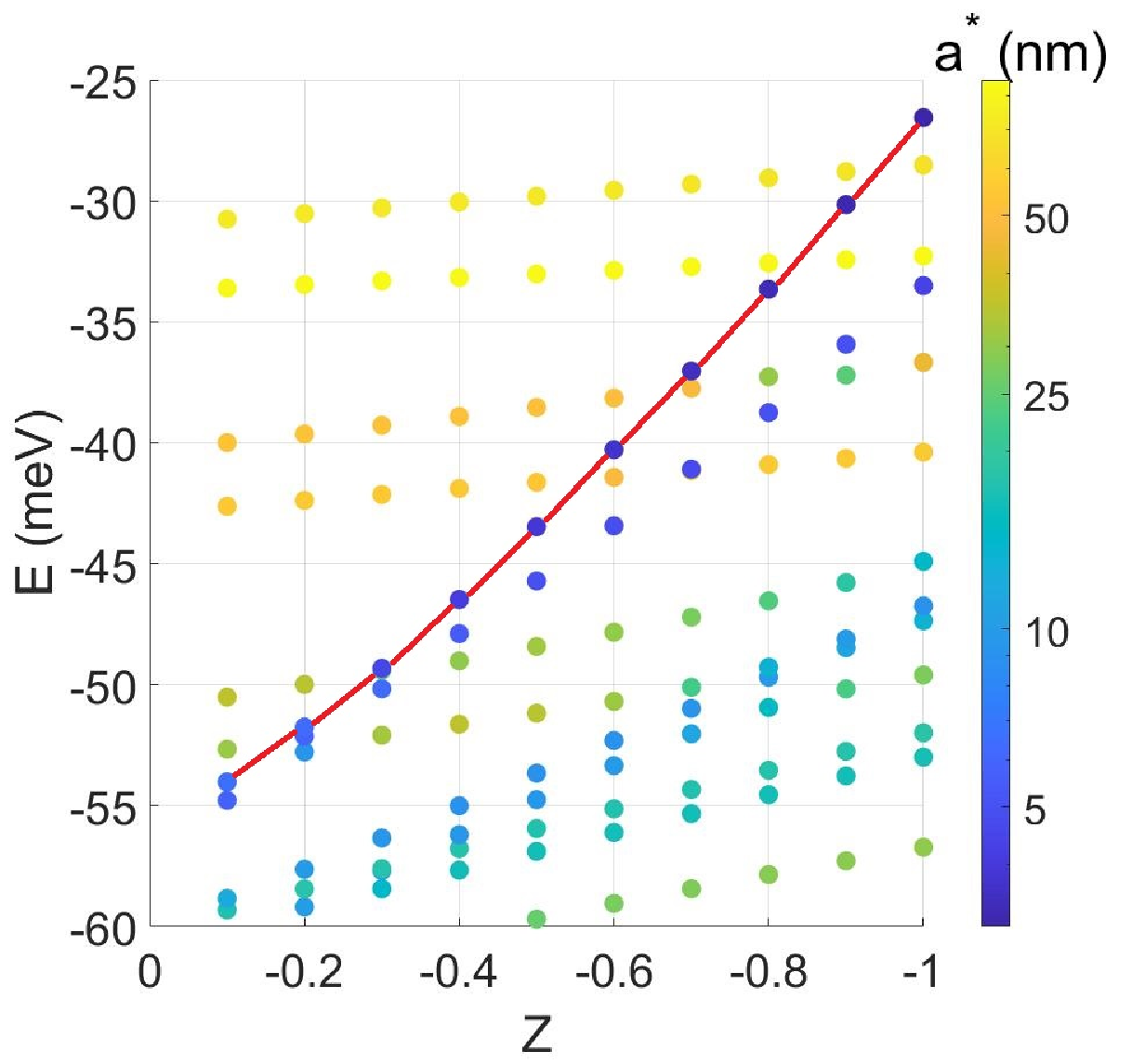}
  \caption{\label{fig:Z} Computed energy levels with respect to the top of the valence band in bulk HgTe for quantum well thickness $d_{\rm{QW}} =6$\,nm and the orbital quantum number $m=0$ as a function of the impurity charge $Ze$ ($e$ -- elementary charge). Red solid line shows the acceptor ground state superimposed on the quantum well states whose energy varies little with $Z$. Colour scale presents the apparent in-plane localization radius $a^*$.}
\end{figure}

An interesting aspect of acceptor states in HgTe and related systems is the co-existence of acceptor
levels and band states in a certain energy range. In the case of bulk samples or wide QWs, the acceptors form resonant states with the conduction band.
In contrast, near the topological phase transition in the 2D case, i.e., in the vicinity of $d_{\mathrm{QW}}=6\,\mathrm{nm}$ in the case of HgTe QW, the acceptor impurity band overlaps with the valence band portion of the Dirac cones \cite{Dietl:2023_PRL}. To tell the band and acceptor states, we have
analyzed the relationship between $E$, $a^*$ and the value of the acceptor charge $Ze$ ($e$ -- elementary charge). Figure \ref{fig:Z} presents the identification of the acceptor states for $d_{\mathrm{QW}}=6\,\mathrm{nm}$, $l_{\mathrm{max}}=10$, $n_{\mathrm{max}}=25$ and $a_1=0.5\,\mathrm{nm}$. Finally, we recall that there is an energy overlap between acceptor levels and topological edge states in topological QWs, which results in a strong Kondo coupling of edge electrons and holes localized on acceptors \cite{Dietl:2023_PRL,Dietl:2023_PRB}.

\begin{table} [htb]
\caption{\label{tab1} Comparison of the acceptor energy levels $E$ (in meV) and its localization radiuses $a^*$ (in nm) obtained using the orthonormal ($E_{\mathrm{ort}}$, $a^*_{\mathrm{ort}}$) and non-orthogonal ($E_{\mathrm{no}}$, $a^*_{\mathrm{no}}$) base functions [formulas (\ref{ort}) and (\ref{no}), respectively] for two QW thicknesses $d_{\mathrm{QW}}$ (in nm) and different values of the quantum number $m$. States corresponding to $m=0$ and $m=-1$ (or $m=-1$ and $m=2$) are degenerate in the presence of the time reversal symmetry -- slight differences between the computed values reflect numerical errors in the coefficients $w_{|m_i|}(k,l)$ that differ for different values of $|m_j|$. The values of $E_{\mathrm{no}}$ and $a^*_{\mathrm{no}}$ have been obtained for $l_{\mathrm{max}}=15$, $n_{\mathrm{max}}=25$.}
\vspace{4pt}
\centering
\begin{tabular}{l l l l l l}
$d_{\mathrm{QW}}$& $m$ & $E_{\mathrm{ort}}$  & $E_{\mathrm{no}}$  & $a^*_{\mathrm{ort}}$  & $a^*_{\mathrm{no}}$  \\
\hline
  8& 0&-12.04   &-11.91   &3.591  &5.802  \\
  8& -1&-11.93 &  &3.660  & \\
  8&  1&-17.68  &  &4.544  & \\
  8& -2&-17.26  &  &4.571   & \\
  6& 0&-26.55  &-26.41  &3.135   &2.538 \\
  6& -1&-26.40 &  &3.221   & \\
  6& 1&-34.46  &  &4.359   & \\
  6& -2&-33.51   &  &4.295 & \\
\hline

\end{tabular}
\end{table}

Table \ref{tab1} presents a comparison of the obtained values of $E$, $a^*$ using the set of the orthonormal wave functions compared to results obtained using wave functions that are not orthogonal \cite{Dietl:2023_PRB}. This comparison shows that the ground state of the acceptor corresponds to $m=0$ and $m=-1$, whose degeneracy is imposed by the time reversal symmetry. As could be expected, the eigenvalues $E$ obtained using both diagonalization methods are almost identical. However, the magnitudes of
the localization radius $a^*$ differ quite significantly, which substantiates the use of the diagonalization method proposed here. We have also calculated the inverse participation numbers along the $z$ axis, obtaining $z^*=6.950\,\mathrm{nm}$ for $d_{\mathrm{QW}}=8\,\mathrm{nm}$ and $z^*=5.530\,\mathrm{nm}$ for $d_{\mathrm{QW}}=6\,\mathrm{nm}$.

Another important information is the fractional contribution $|A_j|^2$ of the particular eight Kohn-Luttinger amplitudes $u_j$ to the acceptor wave functions. For the acceptor ground state with $m =0$ and for $d_{\mathrm{QW}} =8$\,nm, we find $|A_j|^2 =
0.007$,   0.001,   0.285,   0.477, 0.008,   0.221,  0.0002, and 0.0005, which sum up to 1.00. Similarly, for $d_{\mathrm{QW}}=6$\,nm, $|A_j|^2 =
0.009$,  0.002,   0.232,   0.484, 0.016,   0.256,   0.0003,   and 0.0008. As seen, the largest contribution comes from the Kohn-Luttinger amplitudes $p_{3/2,\pm 3/2}$ and $p_{3/2,1/2}$. The time-reversal operator ${\cal{T}}$ serves to obtain the Kramers partner corresponding to $m =-1$, $\Psi^{(m =-1)} = {\cal{T}}\Psi^{(m =0)}$.

 \section{Application to bound magnetic polarons}
 \label{sec:BMP}
 As mentioned in the Introduction, acceptor bound magnetic polarons (BMPs) play a crucial role in magnetic topological systems. In particular, BMPs weaken Kondo scattering of edge electrons by acceptor holes and decouple resonant acceptor levels from band states. According to the analytical solution of the central spin problem, the polaron energy $\epsilon_{\mathrm{p}}$ determines BMP energetics and thermodynamics \cite{Dietl:2015_PRB},
 \begin{equation}
\epsilon_{\mathrm{p}}(T) =
\frac{{\cal{J}}_{sp-d}^2\chi_{\mathrm{TM}}(T)}{(2g\mu_{\mathrm{B}}N_0)^2}\int d\bm{r}|\psi|^4(\bm{r}),
\label{eq:ep}
\end{equation}
where $N_0$ is the cation concentration and the magnitude of the exchange energy ${\cal{J}}_{sp-d}$ is given here by the values of the $s$--$d$ and $p$--$d$ exchange integrals $N_0\alpha$ and $N_0\beta$, respectively,
weighted by orbital content of the acceptor wave functions $A_j$;
$\chi_{\mathrm{TM}}(T)$ is the magnetic susceptibility of magnetic impurities with spin $S$ and the Land\'e factor $g$ in the absence of acceptors; $\psi({\bm{r}})$ is the acceptor wave function, so that the integral is the inverse participation number. A numerical evaluation of this integral gives 0.0021 and 0.0033 nm$^{-3}$ for $d_{\mathrm{QW}} =8$ and 6\,nm, respectively, close to the values of $1/2\pi a^{*2}z^* = 0.0019$ and 0.0029 nm$^{-3}$, respectively. 
Furthermore, the presence of the spin-orbit interaction and axial symmetry means that the polaron energy can depend on the angle $\theta$ between BMP magnetization and the growth direction $z$.

As an example, we consider Hg$_{1-x}$Mn$_x$Te QWs in the absence of an external magnetic field, for which $S = 5/2$, $g = 2.0$, $N_0\alpha = 0.3$\,eV and $N_0\beta = -0.7$\,eV \cite{Autieri:2021_PRB}. Furthermore, we take into account the values quoted in Sec.~\ref{sec:res} for the ground state acceptors residing in the QW centre, i.e., $a^*$ and $z^*$, and the dominant Kohn-Luttinger amplitudes, whose contribution is described by the magnitudes of $A_1$, $A_3$, $A_4$, and $A_6$. To find the value and angular dependence of ${\cal{J}}_{sp-d}$, we determine the expected value of the $sp$-$d$ hamiltonian for various directions $\theta$ and $\phi$ of a unit magnetization vector $\bm{w}$ and the correspondingly oriented, in the Bloch sphere, the eight component hole spinor $\bm{j}$ built of $m=0$ and $m = -1$ basis. For Mn magnetization oriented parallel to $\bm{j}$ we obtain,
\begin{eqnarray}
E_{sp-d} = \frac{1}{4}\{N_0\alpha|A_1|^2[1 + b_1 + \cos(2\theta)(1 - b_1)] + \\ \nonumber
 \beta[(|A_3|^2 - |A_6|^2)(1 +\cos(2\theta)) + \\  \nonumber
 \frac{1}{3}|A_4|^2[1 + 2b_4 + \cos(2\theta)(1 - 2b_4)]]\},
\label{eq:Espd}
\end{eqnarray}
where $b_i = \cos(\mbox{Arg}(A_i^*A_i^*))$. As $A_1$ is real whereas $A_4$ imaginary we have $b_1 =1$ and $b_4 = -1$. Because of the axial symmetry, $E_{sp-d}$ is independent of $\phi$.
% ...
%
%

According to Eq.~\ref{eq:Espd} and the determined values of $A_i$, $|{\cal{J}}_{sp-d} | = 2|E_{sp-d}|$ attains a maximum for $\theta =\pi/2$, i.e., for the in-plane direction of BMP magnetization. Such magnetic anisotropy means that BMP spin-splitting $\Delta$ is enhanced by in-plane components of magnetization fluctuations. The resulting most probable value $\bar{\Delta}$ is given implicitly by \cite{Dietl:2015_PRB},
\begin{equation}
\bar{\Delta}^2-2\bar{\Delta}\epsilon_{\mathrm{p}}(T) \tanh(\frac{\bar{\Delta}}{2k_{\mathrm{B}}T}) - 4\epsilon_{\mathrm{p}}(T)k_{\mathrm{B}}T = 0.
\end{equation}

For paramagnetic Hg$_{1-x}$Mn$_{x}$Te QW, where $x =1.2$\% we find that $\bar{\Delta}$ is larger than $k_{\rm{B}}T$ at $T \le 7.4$ and 8.3\,K for $A_i$ and $b_i$ parameters quoted above for $d_{\mathrm{QW}} = 8$ and 6\,nm, respectively. In those $x$-dependent temperature regimes, the presence of BMP significantly enhances the Coulomb gap and diminishes Kondo scattering. A steady improvement of conductance quantization accuracy on lowering temperature below 4\,K was found experimentally in Hg$_{0.988}$Mn$_{0.012}$Te QW with $d_{\mathrm{QW}} = 9$\,nm and the topological band gap of about 10\,meV \cite{Shamim:2021_NC}. No such recovery of quantization precision at low temperatures has been reported for non-magnetic HgTe QWs. Because of a direct relation between the BMP characteristic energies ($\epsilon_p$,$\bar{\Delta}$) and the impurity localization radiuses ($a^*$ and $z^*$), the explanation of the temperature recovery of quantization accuracy constitutes a worthwhile experimental verification of our theory. We also expect that our work will stimulate optical studies of impurity and BMP energies in magnetic quantum wells.


 \section{Conclusions}
We have proposed to employ a set of orthonormal radial functions for determining wave functions of localized states in semiconductors and semiconductor quantum structures by the $kp$ method. We have demonstrated the applicability of such an approach for a rather demanding case, namely for acceptor states in topological quantum wells, in which impurity energies overlap with a continuum of band states. That example indicates that the proposed method is quite powerful and can be applied for other settings. Having quantitative information on the wave function, we have determined characteristics of bound magnetic polarons that appear in the presence of magnetic ions. Our results substantiate quantitatively a surprising recent conclusion \cite{Dietl:2023_PRL,Dietl:2023_PRB,Sliwa:2023_arXiv,Cuono:2023_arXiv} that doping of quantum wells with paramagnetic impurities stabilizes all three quantum Hall effects by (i) decoupling band and resonant impurity states; (ii) reducing Kondo backscattering of electrons in helical edge state, and also (iii) diminishing parallel hoping conductance of in-gap localized states.

Though we have considered explicitly Mn ions with the open $3d$ shell as the magnetic constituent, the formation of BMP is also expected for topological quantum wells doped with rare earth elements with open $4f$ shells, provided that solubility limits can be overcome. A significant solubility of Eu has already been demonstrated for the case of lead chalcogenides crystallizing in a rock salt structure \cite{Bauer1992MagnetoopticalPO}. However, a large dielectric constant specific to those systems may hamper the presence of bound impurity states.

\section*{Data availability statement}
The numerical code and data supporting the findings of the presented study are available in the Repository for Open Data (RepOD) of the Interdisciplinary Centre for Mathematical and Computational Modelling, University of Warsaw at \href{https://doi.org/10.18150/QYSEFT}{doi:10.18150/QYSEFT}.
\section*{Conflict of interests}
The authors declare no competing interests.
\section*{Orcid IDs}
Dawid Bugajewski: 0009-0002-2327-6037; \\
Tomasz Dietl: 0000-0003-1090-4380.
\section*{Acknowledgements}
This research was partially supported by the “MagTop” project (FENG.02.01-IP.05-0028/23) carried out within the “International Research Agendas” programme of the Foundation for Polish Science co-financed by the European Union under the European Funds for Smart Economy 2021-2027 (FENG).
\section*{References}
\bibliographystyle{iopart-num}
\bibliography{acceptorstates,}
\end{document}